\documentclass[preprint,12pt, authoryear]{elsarticle}

\usepackage{amssymb}
\usepackage{amsmath}
\usepackage{graphicx,epsfig,rotating,color}

\begin{document}

\begin{frontmatter}

\title{How news affect the trading behavior of different categories of investors in a financial market}

\author[1,2,3]{Fabrizio Lillo}
\address[1]{Dipartimento di Fisica, Universit\`a di Palermo, Viale delle Scienze, I-90128, Palermo, Italy}
\address[2]{Scuola Normale Superiore di Pisa, Piazza dei Cavalieri 7, 56126 Pisa, Italy}
\address[3]{Santa Fe Institute, 1399 Hyde Park Road, Santa Fe, NM 87501, USA}

\author[1]{Salvatore Miccich\`e}
\author[4,5]{Michele Tumminello}
\address[4]{Department of Social and Decision Sciences, Carnegie Mellon University, Pittsburgh, PA15213, USA}
\address[5]{Dipartimento di Scienze Statistiche e Matematiche ``Silvio Vianelli", Universit\`a di Palermo, Viale delle Scienze, I-90128, Palermo, Italy}

\author[6]{Jyrki Piilo}
\address[6]{Turku Centre for Quantum Physics, Department of Physics and Astronomy, University of Turku, 
FI-20014 Turun yliopisto, Finland}

\author[1]{Rosario N. Mantegna}

\begin{abstract}
We investigate the trading behavior of a large set of single investors trading the highly liquid Nokia stock over the period 2003-2008 with the aim of determining the relative role of endogenous and exogenous factors that may affect their behavior. As endogenous factors we consider returns and volatility, whereas the exogenous factors we use are the total daily number of news and a semantic variable based on a sentiment analysis of news. Linear regression and partial correlation analysis of data show that different categories of investors are differently correlated to these factors. Governmental and non profit organizations are weakly sensitive to news and returns or volatility, and, typically, they are more correlated with the former than with the latter. Households and companies, on the contrary, are very sensitive to both endogenous and exogenous factors, and volatility and returns are, on average, much more relevant than the number of news and sentiment, respectively. Finally, financial institutions and foreign organizations are intermediate between these two cases, in terms of both the total explanatory power of these factors and their relative importance.    
\end{abstract}


\end{frontmatter}


\section{Introduction}
\label{Introduction}
The efficient market hypothesis assumes that financial markets discount immediately all information available about the listed assets. In general the flux of information is both endogenous and exogenous and this leads to a classification of different forms of efficiency. It is natural therefore that a large empirical literature exists with the aim of measuring the role of exogenous and endogenous sources of information in explaining price dynamics, both in absolute and in relative terms. The seminal work of \cite{Cutler1989} started a stream of research trying to connect exogenous news with price movements. As detailed in the next literature review section, more recent papers have investigated stock price reaction to news, e.g. \cite{Chan2003,Vega2006}, the correlation between high/low pessimism of media and high market trading volume \citep{Tetlock2007}, the relation between the sentiment of news, earnings and return predictability \citep{Tetlock2008}, the correlation between the volume of searching of news in the Google searching engine and many financial indicators of stocks \citep{Da2011}, the role of news in the trading action of short sellers \citep{Engelberg2010}, the role of macroeconomic news in the performance of stock returns \citep{Birz2011}, and the high frequency market reaction to news, \citep{Joulin2008,Gross2011}. All these papers are concerned with the relation between news and price movements. A different and less explored stream of research, to which the present paper belongs, investigates the role of news on the trading and investment decisions of single investors. The main difficulty of this type of research is the availability of micro data about the activity of single investors. Recently, some papers have investigated how news affect the selection of stocks performed by single investors \citep{Barber2008}, the role of individual investor decisions in causing post-earning announcement drift \citep{Hirshleifer2008},  and the relation between high news attention and the level of trading of single investors \citep{Yuan2008}.

Financial markets are extremely heterogeneous systems, and investors represent an important source of heterogeneity. Investors are different in many respects, including their risk profile, the size of their investment, the regulatory constraints to which they are subject, the information they have access to, etc. One clear form of heterogeneity among investors, which incorporates most of the aforementioned differences, is the {\it category} to which an investor belongs. A large financial institution is clearly different from an household, or from a governmental institution and this difference might be reflected in the way each investor reacts to exogenous (news) or endogenous (price returns or volatility) factors. The concept of category is related to a specific classification and it is not unique. In this paper, we will use a classification given by the data we are using, which allows us to discriminate between non-financial corporations, financial and insurance corporations, general governmental organizations, non-profit institutions, households, and foreign organizations. 

More specifically, in the present study, we investigate how different categories of single investors react to exogenous and endogenous factors by looking at their trading activity, both in terms of being active in the market and in terms of the decision to buy or to sell an asset. 
To this aim we make use of two very detailed datasets. We consider the Nokia stock and we have access to a database containing the trading activity of all the investors whose financial ownership is recorded by the Finnish Central Securities Depository.  These data allow us to classify single investors in terms of the category mentioned above, and to characterize the buying and selling activity on Nokia with a daily resolution. For the exogenous news we consider all the Thomson Reuters news released during the time period 2003-2008 and containing information about the company Nokia. 

The number of daily news about Nokia gives us a signal about the intensity of exogenous information without interpreting the content of the news.
In order to have a semantic interpretation of the each news and to classify it in terms of good or bad news we use the General Inquirer (http://www.wjh.harvard.edu/~inquirer/), a well-known content analysis program which is using the General Inquire categories from the Harvard psychosocial dictionary. We construct a simple proxy of the sentiment of news arriving into the market by applying the General Inquirer and measuring the absolute or relative difference between positive and negative words in the headline.  

The main analysis of the paper is a linear regression and partial correlation analysis which allows us to assess for each category of investors the absolute and relative role of endogenous (price return and volatility) and exogenous (number of news and sentiment indicator) factors in explaining the decision to trade and, when they trade, the decision to buy or to sell. As detailed below we find a different behavior among different categories of investors. Compared to other categories, governmental and  non profit institutions are in absolute terms less affected by endogenous and exogenous factors. In relative terms, they are more affected by news than by price dynamics. On the contrary, trading action of households and non financial companies is significantly explained by the regression, but returns and volatility are more important than exogenous news. Finally, we show that financial institutions and foreign institutions display an intermediate behavior, but endogenous factors are more important than exogenous ones.  

The paper is organized as follows: 
In Section II we review the empirical literature on news, price dynamics, and trading activity. In Section III we describe the databases used in our study. In Section IV we introduce the variables used to characterize the trading activity of the investors and the proxies used for the flux of news, volatility and sentiment indicators. Section V presents and discuss the results obtained from the regression analysis, and Section VI concludes. 

\section{Literature review}
 \label{Lr}
 
There is a vast literature about the role of news in financial markets. We can divide the literature in two streams of research. The first and older one considered the problem of how news affect asset price. The second stream of research, more closely related to the present paper, considered the problem on how single investors react to news announcements. This type of analysis has been possible only starting recently because of the availability of large datasets with records of the trading history of single investors. 
In this section we review these two streams of literature.

Many papers have investigated price reaction to news since the pioneering work of \cite{Cutler1989} that estimated for the first time the fraction of the variation that can be attributed to economic news in aggregated stock returns. A few years later, \cite{Ederington1993} studied the impact of scheduled macroeconomic news announcements on interest rate and foreign exchange futures markets. \cite{Engle1993} investigated how new information is incorporated into volatility estimates in the presence of asymmetry in the impact of news. \cite{Mitchell1994} studied the relation between the daily number of Dow Jones news and aggregate measures of market activity such as trading volume and market returns. 

Starting from 2003, new studies using comprehensive databases of news appeared in the literature. \cite{Chan2003} showed that stocks experiencing negative returns concurrent with the arrival of a news story continued to underperform their peers. The same public news database was successfully used by \cite{Vega2006}, together with the estimation of the probability of private information-based trading, to empirically measure the effect of private and public information on post-announcement drift. The sentiment carried by news impacting the market was first investigated by using the daily content from a popular {\it Wall Street Journal} column \citep{Tetlock2007}. In his study, Paul Tetlock found that news with negative sentiment predicts downward pressure on market price followed by a reversion to fundamentals. In a successive study, \cite{Tetlock2008} examined whether quantitative measures of sentiment of the text of news can be used to predict individual firms' accounting earnings and stock returns. In their study they concluded that linguistic content of news captures otherwise hard to quantify aspects of firms' fundamentals that are quickly reflected into stock prices. Another study investigated the role of dissemination of information on security pricing \citep{Fang2009}, showing that stocks with no media coverage earn higher returns than stocks with high media coverage. The role of investors' attention was also considered from the different perspective of information demand in a study of the Google Search Volume Index \citep{Da2011}. In this study, authors related the Google search Volume Index to a sample of Russel 3000 stocks showing that an increase in the Search Volume Index predicts higher stock prices in the next two weeks and an eventual price reversal. 

The role of news in short sales was investigated by \cite{Engelberg2010}, who found that a negative relation between short sales and future returns is order twice larger on news days than on days without a significant flux of news. The relationship is of the order of four times on days with negative news. The analysis of a large electronic database of news allowed to investigate the role of news in high frequency trading on both volatility \citep{Joulin2008} and price formation and book dynamics \citep{Gross2011}. \cite{Joulin2008} found that volatility patterns around market endogenous jumps and around exogenous news are quite different with endogenous jumps followed by increased volatility and news triggering periods of lower than average volatility. The role of high frequency sentiment indicators on future price trends and bid-ask spreads has been studied by considering the high frequency price evolution of twenty stocks traded at the London Stock Exchange  \citep{Gross2011} and the Reuters NewScope Sentiment Engine, which is a pre-processed set of news data and electronic tools analyzing textual information using linguistic pattern recognition algorithms. Recent studies have also considered the role of specific categories of news, such as macroeconomic news \citep{Birz2011}, and the role of investor sentiment on the market's mean-variance trade off \citep{Yu2011}.

The literature on the role of news on trading decision and activity of single investors is more limited. \cite{Barber2008} tested and confirmed the hypothesis that individual investors are net buyers of stocks frequently discussed in the news. They proposed a model of decision making in which individual investors consider primarily those stocks having attention-grabbing qualities and preferential selection among them is exercised only after attention has limited the choice set. The response of individual investors to news was investigated by Yuan in a study that considered the trading and position information of 78,000 households investing in US markets from January 1991 to December 1996 \citep{Yuan2008}. In his study, Yuan showed that the impact of attention is pervasive across market. High attention causes individual investors to reduce stock positions in good times and moderately increase stock positions in bad times. His results also indicate that attention is one source of the cost of monitoring portfolios and that investors sensitive to news are more subject to the disposition effect. Another investigation on the behavior of individual investors in the presence of public news studied the role of individual investors in causing post-earnings announcement drift \citep{Hirshleifer2008}. Authors found that individuals are net buyers after both negative and positive bold earning announcements. 

\section{Data}
\label{Data}
In this paper we investigate the database maintained by the Euroclear Finland (previously Nordic Central Securities Depository Finland). The database is the central register of shareholdings for Finnish stocks and financial assets in the Finnish Central Securities Depository. Practically all major publicly traded Finnish companies have joined the register. The register reports the shareholdings of all Finnish investors and of all foreign investors asking to exercise their vote right. Both retail and institutional investors are included. The database records official ownership of companies and financial assets and the trading records are updated on a daily basis according to the Finnish Book Entry System. The records include all the transactions, executed in worldwide stock exchanges and in other venues, which change the ownership of the assets.

The database classifies investors into six main categories: non-financial corporations, financial and insurance corporations, general governmental organizations, non-profit institutions, households, and foreign organizations. 
The database is collected since January 1, 1995.  In the present study we investigate the market activity of investors trading the Nokia stock, which was, across the years under investigation, either the most capitalized stock or one of the most capitalized stocks in the Finnish stock market. 

While the database contains very detailed information about the Finnish domestic investors, foreign investors can choose to use nominee registration. In this case, the investor's book entry account provider aggregates all the transactions from all of its accounts, and a single nominee register coded identity contains the holdings of many foreign investors. 
This means that our results describe in a detailed way the actions of all the Finnish domestic investors and those foreign investors who do not use nominee registration, while a very small fraction of the coded identities correspond to a large aggregated ownership.

For this reason, in the present study, we consider only the set of single investors trading the Nokia stock during the period of time from January 2, 2003  to December 30, 2008 (a set of $1,510$ daily records) and we investigate all the market transactions performed by them. Single investor means here a retail or an institutional investor that do not use nominee registration (essentially all the Finnish investors). The total number of investors is $141,190$ and the total number of transactions is $7,494,104$. Table \ref{TableDescr1} reports the number of investors, the number of transactions, and the traded volume for the six categories. 

\begin{table*}
\caption{\label{TableDescr1} Summary of the number of investors (\# ids), the number of transactions ($N$), and the exchanged volume ($V$) for the Nokia single investors in the period Jan. 2, 2003 - Dec. 30, 2008.  Volume is given in millions of shares. The investors are divided in the six categories. Nominee registered investors are not considered. Note that for transactions between two single investors the volume is counted twice, once for the buyer and once for the seller.}
\begin{center}
\begin{tabular}{|l|rrr|}
  \hline
  Category				& \# ids &$N$~~~ 	&$V$~~ 	\\
  \hline
Companies			 	&8,396	&	 1,009,226 	& 4,825		\\
 Financial 			 		&392		&	 4,079,174 & 21,402	 	\\ 
 Governamental			&124		&	 39,278	& 1,985		\\
 Non profit					&922		&	 21,778 & 248		\\
 Households				&129,952	&	 1,555,096	& 1,993		\\
 Foreign 					&1,405	&	789,552	 	& 7,685		\\
 \hline
Total						&141,190	&	 7,494,104 & 38,138		\\
\hline
\end{tabular}
\end{center}
\end{table*}

In this paper we investigate the relation between the trading  of Nokia investors, the Nokia price dynamics, and the flux of news about Nokia. As a source of news reaching financial markets worldwide we use the Headlines of the NewsScope archive of news released in English by Thomson Reuters during the investigated time period. Specifically, from the complete NewsScope archive we have extracted all headlines in English language labeled with at least one Nokia Reuters Instrument Code \footnote{The RICs used to extract the headlines  are NOK.W, NOK1V.HE, NOK1V.AS, NOKN.MX, NOKA.BA, NOKy.BE, NOK.MW, NOKy.F, NOK1VEUR.VIp, NOK1VEUR.Ip, NOK1VEUR.STp, NOKS.HA, NOKS.H, NOKS.DE, NOKy.D, NOKAc.BA, NOKy.MU, NOKy.DE, NOK1VM0110.HE, NOK1VEUR.PZ, NOK, NOK1VEUR.DEp, NOKI.ST, NOKS.BE, NOKS.F, NOK.DF, NOK.N, NOKAd.BA, NOK.P, NOK.C, NOK1VEUR.MIp, 0HAF.L, NOK1VEUR.PAp, NOK1V.MI ,NOKS.D, NOKS.MU}. The set comprises 11,484 unique headlines. Each headline is associated with one or more release time (multiple releases of the same headline are frequent). In case of multiple releases of the same headline we use as time of the headline the time of the first release.
\begin{figure}
\begin{center}
\includegraphics[scale=0.45,angle=0]{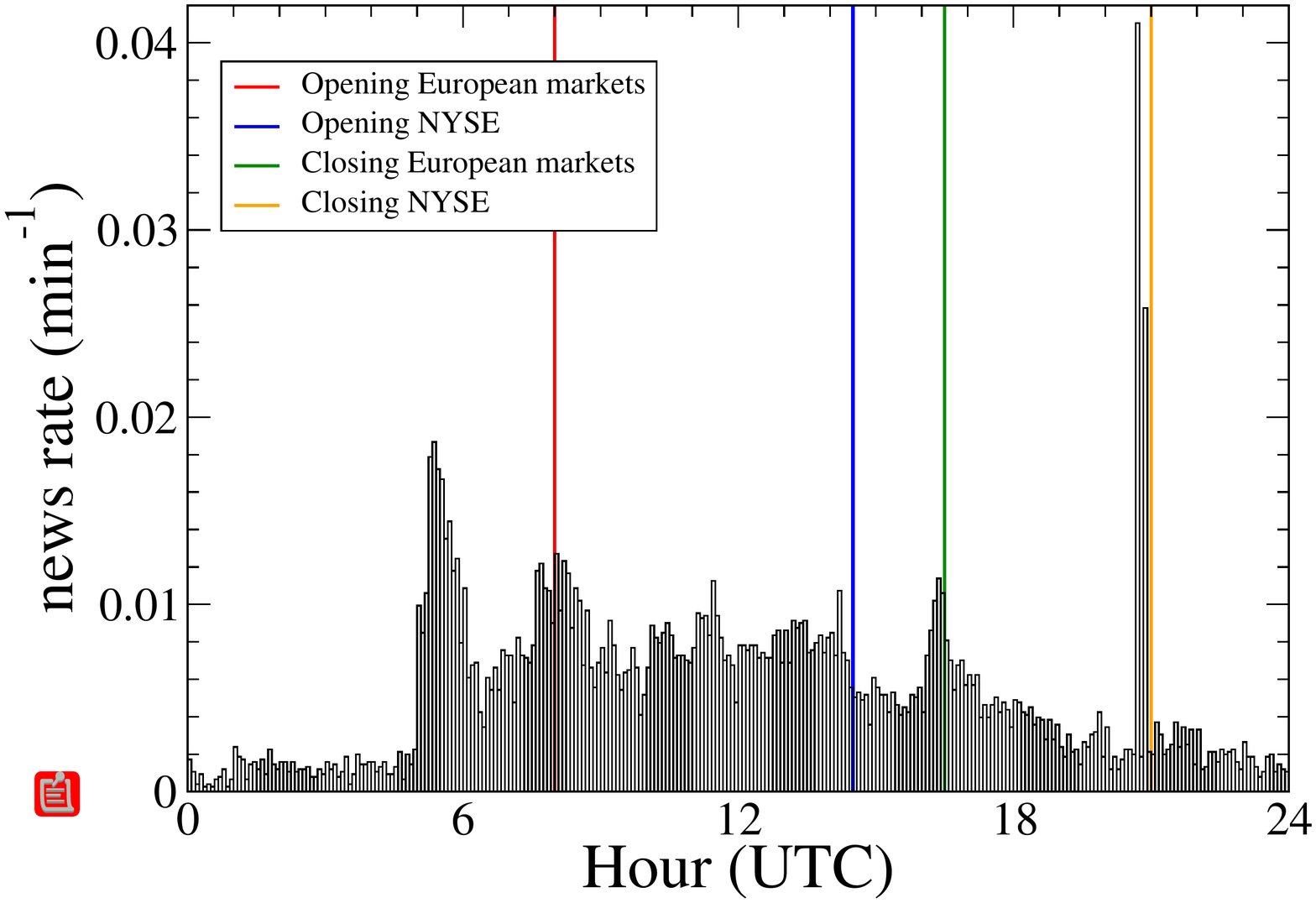}
\caption{Average daily pattern of the arrival rate of news on the Nokia company. The rate is measured in number of headlines per minute. The vertical lines indicate the time of opening and closing of European and New York Stock Exchange market. Data are adjusted for the daylight saving time.}
\label{DFhead} 
\end{center}
\end{figure}
In Fig. \ref{DFhead} we show the average daily pattern of the arrival rate of  Nokia news per minute as a function of the time of the day. Time is computed in coordinated universal time (UTC) and is corrected for the setting of daylight saving time in UK and for the difference between UK and US daylight saving time. The figure shows that news start to arrive at an high rate around 5.00 am (UTC time) and the distribution is roughly flat until 4.30 pm, which is the time of market closing in Europe. Spikes of the probability density function are observed around the time of market opening (8.00 am) and closing (4.30 pm) in Europe and opening (2.30 pm) and closing (9.00 pm) of US NYSE and NASDAQ markets. In the figure, opening and closing of markets are indicated with vertical lines. 

We assume that the largest majority of the Finnish investors are trading in European markets and for this reason, in the present study, we consider only the headlines reaching financial markets during European trading hours (from 8.00 am to 4.30 pm UTC time). Note that a clear peak of news arrival is observed around the closing time of European markets (4.30 pm). Since most of these news are probably telling how the market is closing, we repeated the analysis by removing the last 10 minutes of the trading day, but the results presented below remain essentially unchanged. 

\section{Variables characterizing trading activity, price dynamics, and the flux of news}\label{categorical}
\label{Variables}
In our analysis we consider three sets of variables, one containing variables describing the trading action of the investors, one describing the price dynamics, and one describing the news feed. In this section we define the variables and describe some of their statistical properties.

\subsection{Definition of the variables}

The first set of variables characterize the trading activity of single investors belonging to different categories.
The high degree of heterogeneity of investors in the frequency and volume of trading makes difficult to compare, for example, the activity of an household trading small volumes once every three months with the one of a financial institution that trades every day large volumes. Since we are primarily interested in comparing the impact of news on the daily trading of single investors, we use categorical variables that describes their trading activity. Similarly to what \cite{TumminelloNJP} did, we use the daily categorical variables of Selling investors (S), Buying investors (B) and Buying and Selling investors (BS).  

The classification is obtained as follows: for each investor $i$ and each trading day $t$, we consider the Nokia volume sold $V_s(i,t)$ and the Nokia volume purchased $V_b(i,t)$ by the investor $i$ in that day. This information is then converted into a categorical variable with 3 states: primarily buying \emph{B}, primarily selling \emph{S}, buying and selling approximately closing the position \emph{BS}.  The conversion is done by using the ratio
\begin{equation}
q(i,t)=\frac{V_b(i,t)-V_s(i,t)}{V_b(i,t)+V_s(i,t)}.
\end{equation}
We assign an investor a primarily buying state \emph{B} when $q(i,t)>\theta$,  a primarily selling state \emph{S} when $q(i,t)<-\theta$, and a buying and selling state \emph{BS} when $ -\theta \le q(i,t) \le \theta$  with $V_b(i,t)>0$ and $V_s(i,t)>0$. When $V_b(i,t)=V_s(i,t)=0$ we consider the investor not active on day $t$. According to \cite{TumminelloNJP}, in the present study, we set $\theta=0.01$. Roughly speaking, investors in a buy (sell) state can be seen as acting as net buyers (sellers), while investors in a buy and sell state can be thought as intermediaries or day traders.

We use the categorical variables associated to each investor $i$ for each trading day $t$ to compute the time evolution of the number of investors of a given category $K$ performing a specific trading action (buying, selling or buying and selling). Specifically, $N_B^{K}(t)$ is the number of investors of category $K$ classified as buyers at day $t$, $N_S^{K}(t)$ is the number of investors of category $K$ classified as sellers at day $t$, and $N_{BS}^{K}(t)$ is the number of investors of category $K$ classified as buying and selling at day $t$. 

From these variables we obtain the derived variables
\begin{eqnarray}
N^{K}(t)=N_B^{K}(t)+N_S^{K}(t)+N_{BS}^{K}(t) \\
\Delta N^K_{A} (t)=N_B^{K}(t)-N_S^{K}(t)\\
\Delta N^K_R(t)=\frac{N_B^{K}(t)-N_S^{K}}{N^{K}(t)}
\end{eqnarray}
The variable $N^K(t)$ quantifies the number of trading investors of category $K$ without discriminating the nature of the trading action. The variables $\Delta N_{A}(t)$ and $\Delta N_R (t)$ quantify the polarization of the trading choices of investors of category $K$ towards a buying decision (positive values) or a selling decision (negative values) in absolute and relative terms, respectively. Note that $N^K(t)$ includes all the investors, including those in \emph{BS} state, whereas $\Delta N_{A}(t)$ and the numerator of $\Delta N_R (t)$ are calculated by using only those in \emph{B} and in \emph{S} state. However the denominator of $\Delta N_R (t)$ contains all the active investors.

Market price dynamics is quantified by considering daily return of Nokia stock traded at the Nordic Stock Exchange, i.e.
\begin{equation}
Ret(t)=\log P(t)-\log P(t-1)
\end{equation}
where $P(t)$ is the closing price at day $t$. We also consider a proxy of the daily volatility of Nokia at the Nordic Stock exchange defined as 
\begin{equation}
Vol(t)=2 \frac{P_{max}(t)-P_{min}(t)}{P_{max}(t)+P_{min}(t)}
\end{equation} 
where $P_{max}(t)$ and $P_{min}(t)$ are the highest and lowest price of Nokia at day $t$, respectively.

We finally consider two variables quantifying the flux of news about Nokia arriving at day $t$. First of all, we consider the number $H(t)$ of Nokia headlines released by Thomson Reuters between 8.00 am and 4.30 pm during trading day $t$. This number is a measure of the ``intensity" of news reaching the market in a given day. However headlines news and the associated stories may bear positive, negative or doubtful information. In analogy to \cite{Tetlock2008}, we quantify the sentiment carried by the headlines by constructing a sentiment proxy using the number of positive and negative words present in them. Positive and negative words are detected by using the General Inquirer (http://www.wjh.harvard.edu/~inquirer/), a well-known content analysis program which is using the General Inquire categories from the Harvard psychosocial dictionary. Once the number of positive ($G(t)$) and negative ($B(t)$) words contained in all the headline news at day $t$ have been computed, we determine the variables
\begin{equation}
S_{A} (t)= G(t)-B(t)~~~~~~~~S_R (t)=\frac{G(t)-B(t)}{G(t)+B(t)}
\end{equation}
giving the absolute and relative, respectively, sentiment of the news in a given day. To avoid spurious discretization effects in the calculation of $S_R (t) \in[-1,1]$ we require $G(t)+B(t)\geqslant5$ to compute the sentiment indicator. When $G(t)+B(t)<5$ we set $S_R (t)=0$. The time evolution of $S_R (t)$ is therefore different from zero only when a significant number of positive and negative words are detected in the headlines. 
 
 \subsection{Descriptive analysis}
 \begin{sidewaystable}
\caption{Summary statistics of the investigated variables.}
\begin{tabular}{l|l||rrrrrrr}
\hline
Variable& Category& Min&	5\% quant.&	95\% quant.&	 Max&	Mean&	Median&	Std Dev\\
\hline
$N^K$    &Companies	  &0	&25	&171.55	&630	&73.96	&59	&57.3\\
            &Financial	  &4	&14	&31	&69	&20.29	&19	&5.798\\
            &Governmental &0 	&1	&12	&32	&4.713	&4	&3.769\\
            &Non profit	   &0	&1	&14	&43	&5.278	&4	&4.883\\
           &Households	   &0	&202	&1234.65	&4619	&538	&424	&423\\
           &Foreign	            &0	&3	&16	&47	&8.356	&7	&4.443\\
\hline
$\Delta N_A^K$ &Companies	   &-332	&-64.55	&83	&564	&4.826	&1	&49.8 \\
            &Financial	   & -25	&-11	         &10	& 45          &0.1576	&0	&6.835\\
            &Governmental &-23	&-6	         &7	& 32	         & 0.1781	&0	&4.127\\
            &Non profit	   &-35	&-8.55	&6	&25	         &-0.9609	&-1	&4.648\\
            &Households	   &-2143	&-494.05	&662.55	&4147	&26.5	&-15	&406.6\\
            &Foreign	            &-43	&-8	         &6	&28	          & -0.7152 &-1	&4.629\\
\hline
$\Delta N_R^K$& Companies &-0.8977	&-0.6475	&0.6917	&0.9224	&0.03161	&0.03252	&0.423\\
                         &Financial	    &-0.8125	&-0.4783	&0.5	&0.8947	&0.009647	&0	&0.3012\\
                      &Governmental &-1	&-1	&1	&1	&-0.003318	&0	&0.6574\\
                      &Non profit	      &-1	&-1	&1	&1	&-0.1573	&-0.2	&0.6798\\
                     &Households	      &	-1	&-0.7428	&0.7164	&0.913	&-0.02797	&-0.05142	&0.4637\\
                     &Foreign	      &-1	&-0.7778	&0.6667	&1	&-0.08613	&-0.1181	&0.4468\\
\hline
\hline
$Ret$	&& -0.1843	&-0.0365	&0.0347	&0.1361	&-0.000206	&0	&0.0234\\
$Vol$  &&	0.0050	&0.01007	&0.0609	&0.2296	&0.02637	&0.0207	&0.02041\\
\hline
\hline
$H$   &&	0	&0	&11	&46	&3.742	&3	&4.449\\
$S_A$&&	-12	&-2.55	&4	&23	&0.492	&0	&2.512\\
$S_R$&&	-1	&-0.2	&0.4286	&1	&0.01795	&0	&0.2172\\
\hline
\end{tabular}
\label{summarystat}
 \end{sidewaystable}

 \begin{figure}
\begin{center}
\includegraphics[scale=0.5,angle=0]{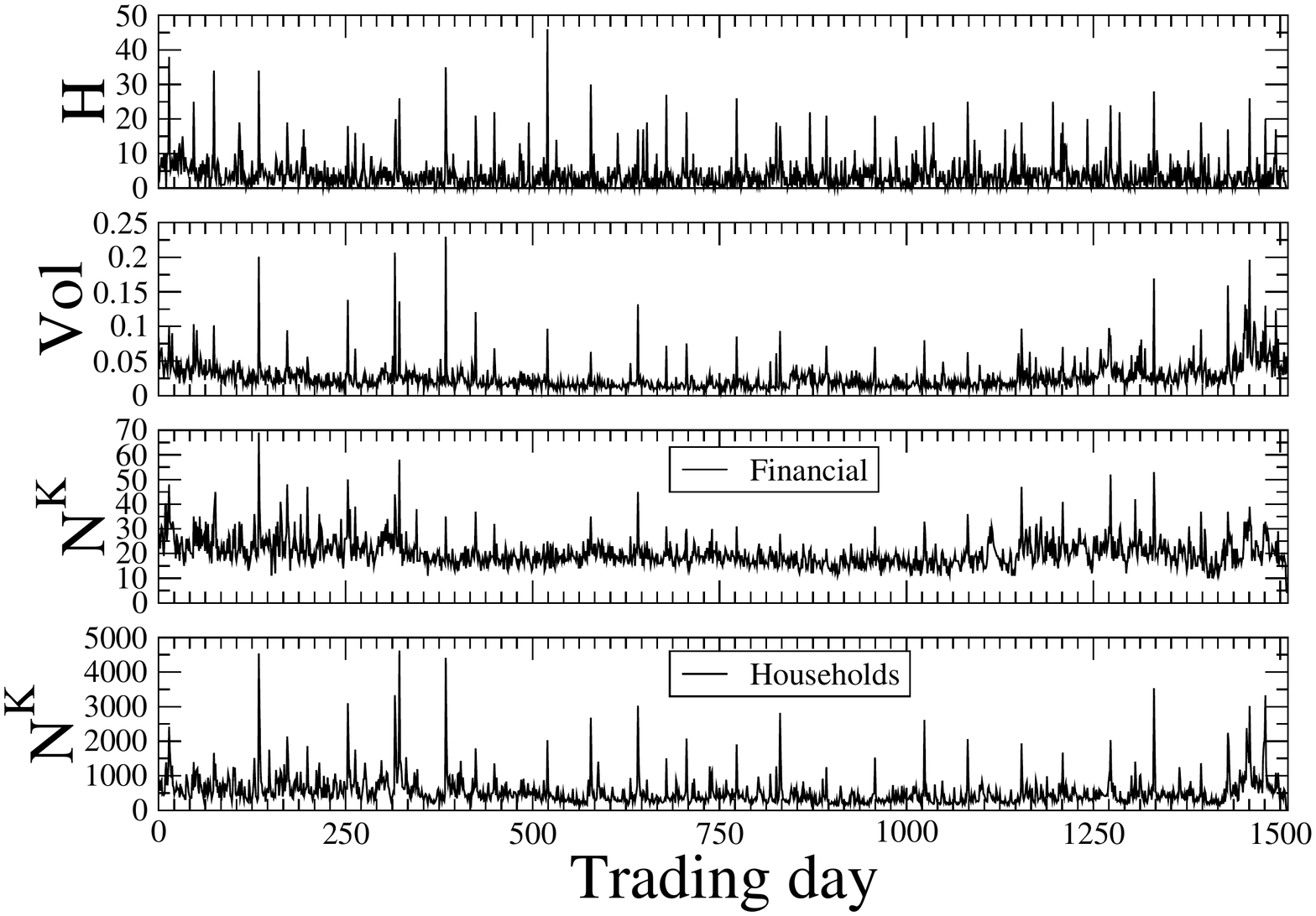}
\caption{From top to bottom the figure shows the time series of the number of Nokia headlines $H(t)$, the daily volatility $Vol(t)$ of Nokia stock, and the time series of $N^K(t)$ for the category of Financial investors and for the category of Households investors.}
\label{ExNa} 
\end{center}
\end{figure}

Table \ref{summarystat} shows the summary statistics of the investigated variables.  We note that for the number of active investors $N^K$ the mean is always larger than the median, indicating a positive skew of the distributions. Moreover the standard deviations are typically quite large. The Table shows that for the governmental,  non profit, and foreign categories, the relative imbalance $\Delta N^K_R$ reaches the minimum and maximum value of $-1$ and $+1$, respectively. 
The presence of these values indicates that, in some days, all the active investors of the considered category took the same market position. However, it is worth noting that the three categories presenting this behavior are the ones having an average number of active investors lower than ten (4.713, 5.278 and 8.356 for governmental, non profit and foreign category respectively). In other words the complete market polarization in most cases involves a limited number of single investors. The same is valid for the occurrence of the minimum value $-1$ observed for households category. This occurrence happened only once during a market session that involved only four households investors.

Figure \ref{ExNa} shows the time series of the number of Nokia headlines $H(t)$, the daily volatility $Vol(t)$ of Nokia stock, and the time series of $N^K(t)$ for the category of Financial investors and for the category of Households investors.  We note that the time series of $H(t)$ presents a background around 4 headlines per day and a series of spikes jumping to 20 headlines or more per day. The flux of news is not clustered in time, in fact the autocorrelation function is significant only at one lag.  The time evolution of the volatility proxy $Vol(t)$ shows that also this quantity is characterized by a typical value (of the order of 2 percent) and by days of huge swings with values of $Vol(t)$ of the order of $20\%$. The spikes of news and volatility are clearly correlated with the spikes of $N^K(t)$ for all the categories of investors. In the figure we show the time evolution of the number of financial and households investors. The overall behavior of the other categories is similar. It is worth noticing that the autocorrelation of $N^K$ is quite persistent and it is statistically significant at $2\sigma$ for more than 30 days.

\begin{figure}
\begin{center}
\includegraphics[scale=0.5,angle=0]{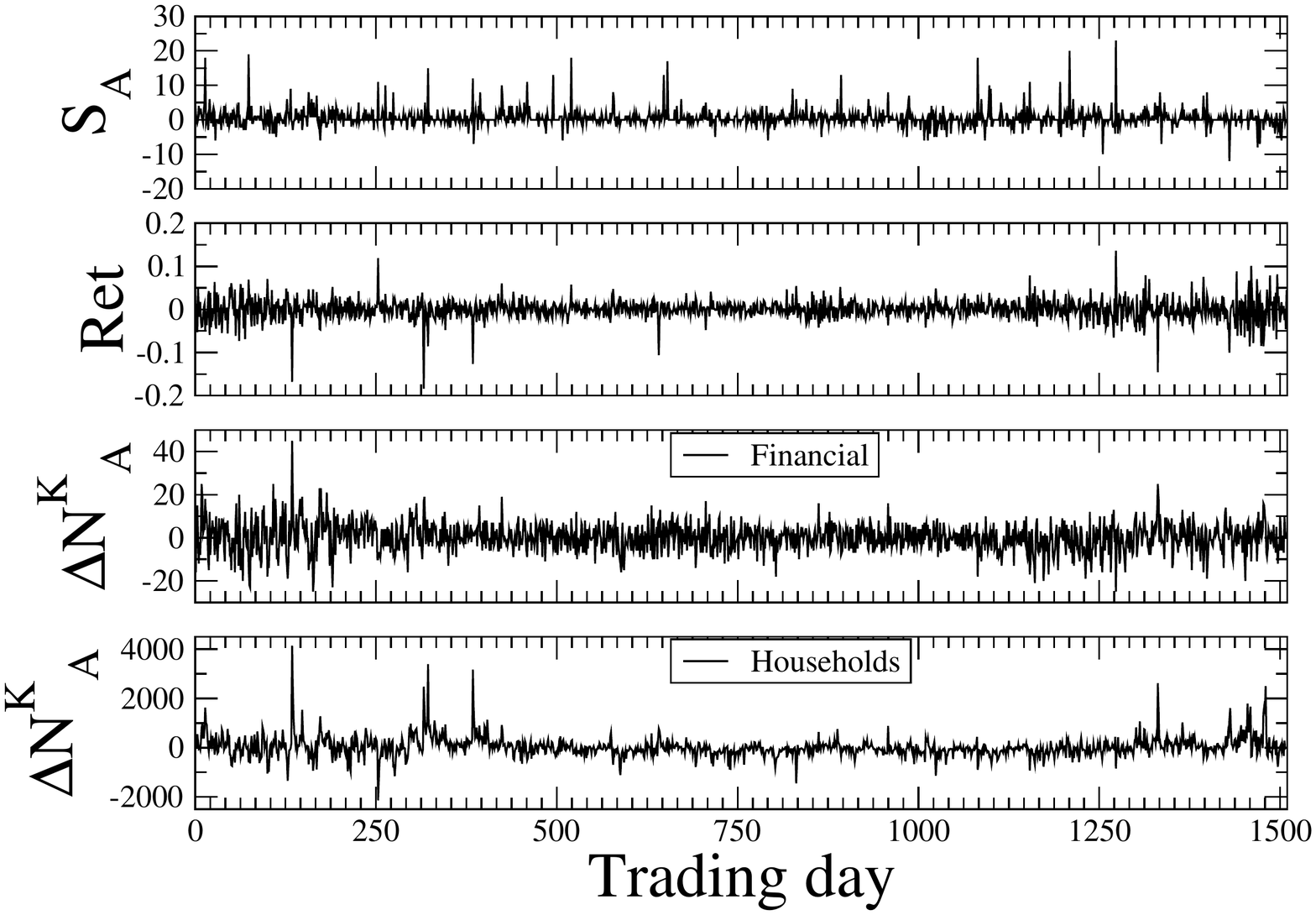}
\caption{From top to bottom the figure shows the time series of the difference $S_{A} (t)$ between the number of positive and negative words in Nokia headlines, of the daily return $Ret(t)$ of Nokia stock, and of  $\Delta N^K_{A} (t)$ for the category of Financial investors and of Households investors.}
\label{ExNbNs} 
\end{center}
\end{figure}

In Fig. \ref{ExNbNs} we show the time series of the difference $S_{A} (t)$ between the number of positive and negative words in Nokia headlines, of the daily return $Ret(t)$ of Nokia stock, and of  $\Delta N^K_{A} (t)$ for the category of Financial investors and of Households investors.  The time evolution of $S_{A}(t)$ fluctuates around zero but also presents a series of positive and negative spikes jumping to the level of order 20 for positive or 10 for negative words per day.  The time evolution of Nokia return is characterized by a non Gaussian profile of the return probability density function and volatility clustering.   
Spikes of $\Delta N^K_{A} (t)$ are also detected but they are in general less pronounced than in the case of $N^K(t)$ (see the bottom two panels of Fig. \ref{ExNa}) suggesting that the interpretation of news and/or endogenously extracted market information is usually different among investors of the same category. However, some pronounced spikes are still observed showing that in some occasion investors' categories take the same kind of trading action. Moreover autocorrelation analysis shows that for financial institutions $\Delta N^K_A$ is persistent only at one lag, while for households the autocorrelation function is statistical significant up to 30 days. This fact indicates the presence of persistent "moods" the household investors are following in their trading actions.
 
\begin{figure}
\begin{center}
\includegraphics[scale=0.5,angle=0]{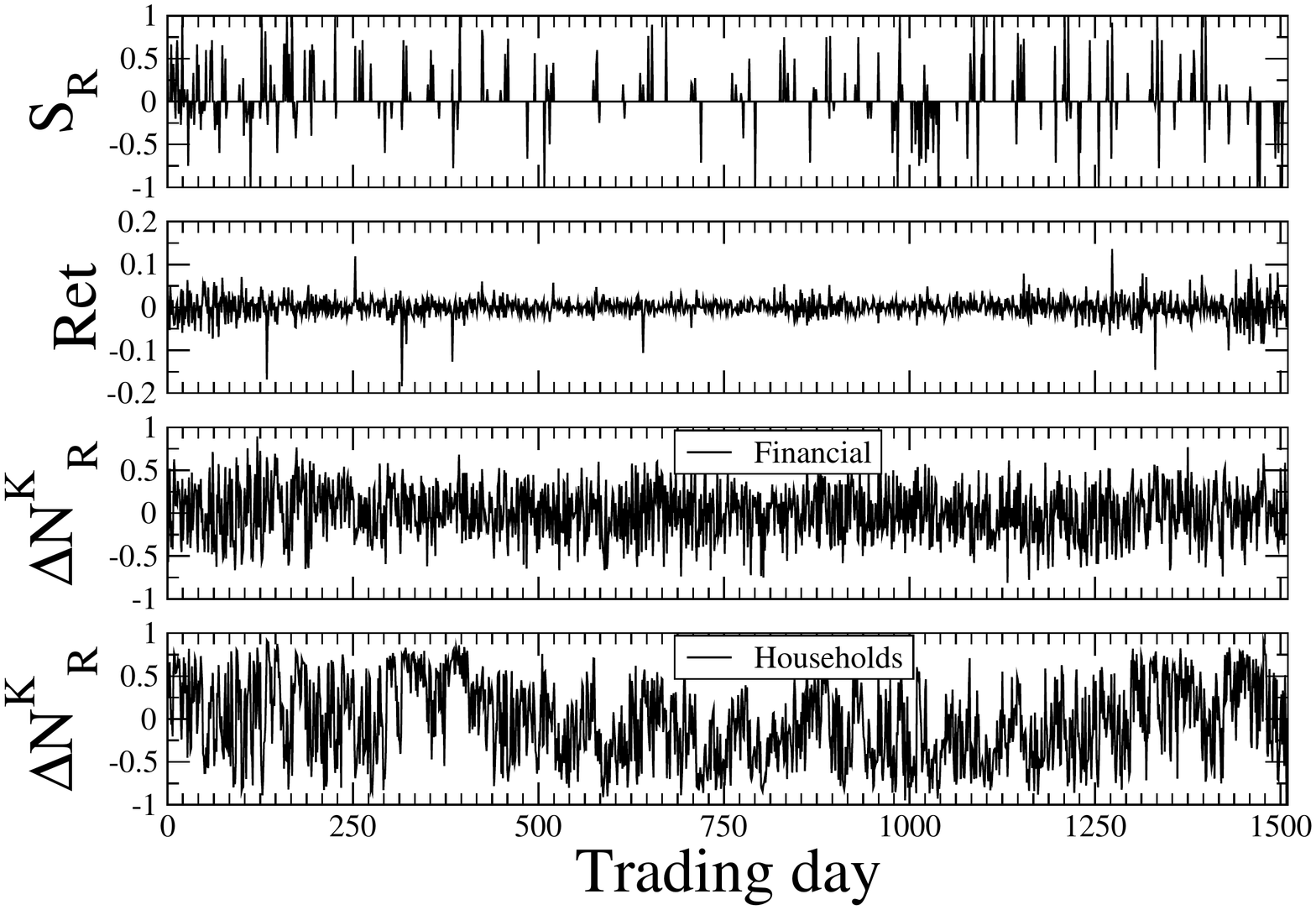}
\caption{From top to bottom the figure shows the time series of the relative sentiment indicator  $S_R (t)$, the daily return $Ret(t)$ of Nokia stock, and the time series of $\Delta N^K_R (t)$ for the category of Financial investors and of Households investors. }
\label{ExDN} 
\end{center}
\end{figure}

In order to investigate the difference between absolute and relative variables we show in Fig. \ref{ExDN} the time series of the relative sentiment indicator  $S_R (t)$, the daily return $Ret(t)$ of Nokia stock, and the time series of $\Delta N^K_R (t)$ for the category of Financial investors and of Households investors.   The time series of $S_R (t)$ presents a series of positive and negative spikes often clustered in time and value.  Quite interestingly, the time series of $\Delta N^K_R (t)$ does not present spikes but rather a noisy oscillation around zero. As in the case of $\Delta N_A^K$, the time evolution of households shows a slow dynamics lasting up to several trading months.

\section{Regression results}
\label{Results}
In this Section we use regression and partial correlation analysis to assess the role of endogenous or exogenous factors in determining the trading behavior of single investors. More specifically, we consider first of all the decision to trade (irrespectively of the specific position taken) and we regress it against an endogenous factor, namely volatility, and against an exogenous factor, namely the number of news. Similarly, in a second step we consider the imbalance (absolute or relative) between buyers and sellers and we regress it against contemporaneous return and the sentiment indicator. Also in this case, the first regressor can be considered endogenous and the second one as exogenous. Before presenting the results of the regression, two comments are in order. First of all, the distinction between exogenous and endogenous is not clear cut and in fact, as we will see, in both cases the regressors are not independent. Second, we shall consider contemporaneous variables and therefore it should be clear that no causality can be attached to our results. To be more explicit, the fact that we find a significant relation in the regression between number of investors and volatility does not necessarily imply that volatility triggers people to trade, but it can be the other way around, i.e. a large number of investors increases the volatility through their trading. Another possibility is that the two variables are influenced by a third unobserved factor.

\subsection{Volatility and  number of news}

Let us consider first how the decision to trade (irrespectively on being a buyer or a seller) is related with volatility $Vol$ and the flux of news $H$ for different categories of investors. Volatility and news are different sources of information, one primarily endogenous and one primarily exogenous to the market, but they are not mutually independent. In fact the empirical correlation between the two variables is Corr$[H,Vol]=0.501$. To interpret the role of the two variables in trading, we consider the linear model 
\begin{equation}
\label{linearNa}
\widehat{N}^K (t)=\alpha_{H} \widehat{H} (t)+ \alpha_{Vol} \widehat{Vol} (t)+\epsilon(t)
\end{equation}
where $\widehat{N}^K$, $\widehat{H}$, and $\widehat{Vol}$ are standardized versions with zero mean and unit variance of $N^K$, $H$, and $Vol$, respectively.

 \begin{sidewaystable}
\caption{Summary of the results of the linear regression of Eq. \ref{linearNa} of the number $N^K$  of trading investors versus the news intensity signal $H$ and the volatility proxy $Vol$. The number in parentheses are the 5\%-95\% confidence intervals under Gaussian hypothesis and by using bootstrap analysis. The last two columns show the results of the partial correlation analysis.}
\begin{tabular}{|l|c|c|c|c|c|}
\hline
Investor& $\alpha_{H}$&$\alpha_{Vol}$&\% variance& $\rho(N^K,H|Vol)$ &$\rho(N^K,Vol|H)$  \\
category&~&~&of residual of $N^K$&&\\
\hline
Companies&0.271 (0.229,0.313)&0.517 (0.475,0.559)&51.8 \%&0.309&0.534\\
bootstrap&~~~~~~~ (0.205,0.335)&~~~~~~~ (0.437,0.597)&~~~~&&\\
\hline
Financial&0.195 (0.149,0.242)&0.479 (0.433,0.526)&63.8 \%&0.207&0.461\\
bootstrap&~~~~~~~ (0.125,0.264)&~~~~~~~ (0.407,0.558)&~~~~&&\\
\hline
Governmental&0.238 (0.183,0.292)&0.192 (0.138,0.246)&86.0 \%&0.215&0.180\\
bootstrap&~~~~~~~ (0.164,0.303)&~~~~~~~ (0.119,0.262)&~~~~&&\\
\hline
Non profit&0.319 (0.269,0.369)&0.270 (0.220,0.320)&73.9 \%&0.305&0.264\\
bootstrap&~~~~~~~ (0.249,0.394)&~~~~~~~ (0.199,0.344)&~~~~&&\\
\hline
Households&0.226 (0.188,0.263)&0.627 (0.589,0.664)&41.4 \%&0.289&0.651\\
bootstrap&~~~~~~~ (0.165,0.285)&~~~~~~~ (0.554,0.697)&~~~~&&\\
\hline
Foreign org.&0.158 (0.109,0.207)&0.442 (0.393,0.492)&70.9 \%&0.160&0.416\\
bootstrap&~~~~~~~ (0.094,0.224)&~~~~~~~ (0.374,0.517)&~~~~&&\\
\hline
\hline
\end{tabular}
\label{NaCtg}
 \end{sidewaystable}

In Table \ref{NaCtg} we show the values of the $\alpha_{H}$ and $\alpha_{Vol}$ coefficients together with the variance of the residual obtained by ordinary least squares. In the Table we also report the 5\%-95\% confidence interval of each coefficient under two null hypotheses. The first is the customary assumption of Gaussian errors, while the second ones are obtained by bootstrapping the data and therefore taking into account the distributional properties of the data. We show the results for each category of investors separately. All the regression coefficients are statistically not consistent with zero (at the given confidence level).
The $\alpha_{H}$ coefficient is ranging from a minimum value of $0.158$, which is observed for the foreign organizations, to a maximum value of $0.319$, which is observed for the non profit organizations. The $\alpha_{Vol}$ coefficient is ranging from a minimum value of $0.192$, which is observed for the governmental organizations to a maximum value of $0.627$, which is observed for households. The variance of the residuals of $\widehat{N}^K$ is ranging from 41.4\% (households) to 86\% (governmental organizations) indicating that in most cases the explanatory value of the two variables is quite significant. 
  
For four categories of investors (companies, financial institutions, households and foreign organizations), the $\alpha_{Vol}$ coefficient is higher than $\alpha_{H}$ indicating that for these investors the market endogenous information has on average an higher explanatory role than market exogenous information conveyed by news on their decision to trade. The case of governmental organizations and non profit organizations is different. For these categories $\alpha_{H}$ is greater than $\alpha_{Vol}$ but the two values are within the 5\%-95\% intervals of the estimated coefficients. Moreover, for these categories the variance of the residual assumes the maximal values which are observed being 86.0 \% for governmental organizations and 73.9 \% for non-profit organizations.  

We complement the regression analysis by computing the partial correlation coefficients of the three variables $N^K$, $Vol$, and $H$. The partial correlation coefficient $\rho(x,y|z)$  between variables $x$ and $y$ conditioning on the variable $z$ is the Pearson correlation coefficient between the residuals of $x$ and $y$ that are uncorrelated with $z$. Partial correlation is clearly related to linear regression. However, the information one obtains from the analysis of regression coefficients is not identical to the one obtained by considering partial correlation. In fact, for normalized variables, the best linear fit $z=\alpha_x x+\alpha_y y+ \epsilon$ gives
\begin{equation}
\label{lineartopartial}
\frac{\alpha_x}{\alpha_y}=\frac{\rho(z,x|y)}{\rho(z,y|x)}\sqrt{\frac{1-\rho^2_{y,z}}{1-\rho^2_{x,z}}}
\end{equation}
where $\rho_{x,z}$ and $\rho_{y,z}$ are the correlation coefficients between $x$ and $z$ and between $y$ and $z$, respectively (please refer to Appendix A fort more details). Only when  $\rho_{x,z}=\rho_{y,z}$ the ratio between the regression coefficients is equal to the ratio between the partial correlations.

Table \ref{NaCtg} shows the values of $\rho(N^K,H|Vol)$ and $\rho(N^K,Vol|H)$ for the six categories of investors. Consistently  with the previous results, we notice that $\rho(N^K,H|Vol)>\rho(N^K,Vol|H)$ for Governmental and Non profit organization, while the reverse is true for the other categories of investors. 

In conclusion, volatility and the flux of news are correlated with the decision to trade of single investors and their relative role, when properly disentangled, is different for different categories of investors. Governmental and non profit institutions are the categories for which news and volatility give less explanatory power of their presence in the market. Moreover, these investors are more sensitive to news than to volatility. Households and companies are much more sensitive to volatility than to news and the variance of their activity explained by these factors is quite high. Also for financial institutions, volatility is more important than news, but the variance is relatively smaller. Finally, foreign organizations are more affected by volatility than by news, but the variance of the regression is quite small.   

\subsection{Returns and sentiment indicator}

By having verified that news play an important role in the decisions of single investors to trade, we now focus our attention on the impact of the sentiment carried by news on Nokia return and on the trading behavior of the single investor to buy or sell a certain amount of Nokia stock. As sentiment indicators we consider both the absolute sentiment indicator $S_{A}(t)$ and the relative sentiment indicator $S_R(t)$. Both daily sentiment indicators are correlated with the daily Nokia return of the Nordic Stock Exchange. The values of the correlation are Corr$[S_{A},Ret]=0.155$ and  Corr$[S_R,Ret]=0.118$. These values are small but statistically significant. In fact the average correlation observed for shuffled time series of $S_{A} (t)$ and $S_R (t)$ with Nokia return is smaller than $0.002$ with a standard deviation of $0.026$.

 \begin{sidewaystable}
\caption{Summary of the results of the linear regression of Eq. \ref{linearDNa} of the difference $\Delta N^K_{A}$   between buying and selling investors versus the absolute sentiment indicator $S_A$ and the stock return $Ret$. The number in parentheses are the 5\%-95\% confidence intervals under Gaussian hypothesis and by using bootstrap analysis. The last two columns show the results of the partial correlation analysis.}
\begin{tabular}{|l|c|c|c|c|c|}
\hline
Investor& $\alpha_{S_{A}}$&$\alpha_{Ret}$&\% variance of & $\rho(N^K_A,S_A|Ret)$ &$\rho(N^K_A,Ret|S_A)$\\
category&~&~&residual of $\Delta N^K_{A}$&&\\
\hline
Companies&0.041 (0.002,0.080)&-0.653 (-0.692,-0.614)&58.0 \%&0.0528&-0.6463\\
bootstrap&~~~~~~~ (-0.031,0.102)&~~~~~~~ (-0.749,-0.572)&~~~~&&\\
\hline
Financial&0.008 (-0.034,0.050)&-0.576 (-0.618,-0.534)&66.9 \%&0.0095&-0.5709\\
bootstrap&~~~~~~~ (-0.055,0.069)&~~~~~~~ (-0.651,-0.516)&~~~~&&\\
\hline
Governmental&-0.019 (-0.070,0.031)&-0.196 (-0.246,-0.146)&95.9 \%&-0.0196&-0.1940\\
bootstrap&~~~~~~~ (-0.083,0.047)&~~~~~~~ (-0.270,-0.125)&~~~~&&\\
\hline
Non profit&-0.026 (-0.076,0.024)&-0.220 (-0.270,-0.170)&94.8 \%&-0.0263&-0.2178\\
bootstrap&~~~~~~~ (-0.116,0.056)&~~~~~~~ (-0.303,-0.136)&~~~~&&\\
\hline
Households&0.060 (0.021,0.099)&-0.655 (-0.694,-0.616)&57.9 \%&0.0781&-0.6474\\
bootstrap&~~~~~~~ (-0.012,0.131)&~~~~~~~ (-0.745,-0.568)&~~~~&&\\
\hline
Foreign org.&-0.006 (-0.050,0.038)&-0.504 (-0.548,-0.460)&74.4 \%&-0.0072&-0.4999\\
bootstrap&~~~~~~~ (-0.072,0.058)&~~~~~~~ (-0.571,-0.438)&~~~~&&\\
\hline
\hline
\end{tabular}
\label{DNAbCtg}
 \end{sidewaystable}

\begin{sidewaystable}
\caption{Summary of the results of the linear regression of Eq. \ref{linearDNR} of the relative difference $\Delta N^K_{R}$   between buying and selling investors versus the relative sentiment indicator $S_R$ and the stock return $Ret$. The number in parentheses are the 5\%-95\% confidence intervals under Gaussian hypothesis and by using bootstrap analysis. The last two columns show the results of the partial correlation analysis.}
\begin{tabular}{|l|c|c|c|c|c|}
\hline
Investor& $\alpha_{S_R}$&$\alpha_{Ret}$&\%  variance of & $\rho(N^K_R,S_R|Ret)$ &$\rho(N^K_R,Ret|S_R)$\\
category&~&~&residual of $\Delta N^K_R$&&\\
\hline
Companies&0.055 (0.014,0.095)&-0.610 (-0.650,-0.569)&63.3 \%&0.0685&-0.6056\\
bootstrap&~~~~~~~ (0.015,0.100)&~~~~~~~ (-0.685,-0.548)&~~~~&&\\
\hline
Financial&0.018 (-0.025,0.062)&-0.520 (-0.564,-0.477)&73.1 \%&0.0212&-0.5170\\
bootstrap&~~~~~~~ (-0.030,0.064)&~~~~~~~ (-0.587,-0.463)&~~~~&&\\
\hline
Governmental&0.021 (-0.029,0.071)&-0.179 (-0.230,-0.129)&96.8 \%&0.0215&-0.1782\\
bootstrap&~~~~~~~ (-0.027,0.075)&~~~~~~~ (-0.225,-0.136)&~~~~&&\\
\hline
Non profit&0.025 (-0.025,0.075)&-0.175 (-0.225,-0.125)&96.9 \%&0.0256&-0.1738\\
bootstrap&~~~~~~~ (-0.028,0.079)&~~~~~~~ (-0.227,-0.130)&~~~~&&\\
\hline
Households&0.068 (0.026,0.110)&-0.565 (-0.608,-0.523)&68.4 \%&0.0811&-0.5615\\
bootstrap&~~~~~~~ (0.025,0.111)&~~~~~~~ (-0.629,-0.512)&~~~~&&\\
\hline
Foreign org.&0.030 (-0.017,0.077)&-0.400 (-0.446,-0.353)&84.2 \%&0.0323&-0.3970\\
bootstrap&~~~~~~~ (-0.015,0.076)&~~~~~~~ (-0.449,-0.354)&~~~~&&\\
\hline
\hline
\end{tabular}
\label{DNRCtg}
\end{sidewaystable}

We analyze the explanatory role of $S_{A}$ and of the Nokia return $Ret$ by considering the linear model 
\begin{equation}
\label{linearDNa}
\widehat{\Delta}N^K_{A} (t)=\alpha_{S_{A}} \widehat{S}_{A} (t)+ \alpha_{Ret} \widehat{R}et (t)+\epsilon(t)
\end{equation}
where $\widehat{\Delta}N^K_{A}$, $\widehat{S}_{A}$, and $\widehat{R}et$ are standardized versions with zero mean and unitary variance of $\Delta N^K_{A}$, $S_{A}$, and $Ret$, respectively.

In Table \ref{DNAbCtg} we show the values of the $\alpha_{S_{A}}$ and $\alpha_{Ret}$ coefficients together with the variance of the residuals obtained by ordinary least squares. As in the previous case, we also report the 5\%- 95\% confidence interval of each coefficient under Gaussian hypothesis and bootstrap analysis and the results are shown for each category of investors separately.
Most of the $\alpha_{S_{A}}$ coefficients are consistent with zero within the 5\%-95\% confidence interval. In fact, for bootstrap confidence intervals all the $\alpha_{S_{A}}$ are consistent with zero, while under normality hypothesis values (slightly) different from zero are observed in the cases of companies and households investors. 
The $\alpha_{Ret}$ coefficient is always statistically significant and it is ranging from a minimum value of $-0.653$ observed for households to a maximum value of $-0.196$ observed for governmental organizations. The variance of the residual of $\widehat{\Delta}N^K_{A}$ is ranging from 57.9\% (households) to 95.9\% (governmental organizations) indicating that still for some of the categories (companies, financial institutions, households and foreign organizations) the explanatory value of the two variables, and especially of the return, is significant. 

The $\alpha_{Ret}$ coefficients are in several cases (companies, financial institutions, households and foreign organizations) negative large values indicating that, for these categories, the market polarization of trading actions is strongly anticorrelated with the Nokia return. The majority of single investors of these categories are therefore buying when the Nokia price goes down and selling when the price goes up. This is reminiscent of a contrarian behavior, even if in this case we are considering contemporaneous rather than lagged correlation between return and decision to buy or sell. 
A similar behavior is also  observed, although to a less pronounced level, for governmental and non-profit organizations investors. 
The partial correlations showed in Table  \ref{DNAbCtg} confirm this results (see below for an extended  comment) . 

These results are essentially confirmed by the analysis concerning the explanatory role of the relative sentiment indicator $S_{R}$ and of the Nokia return $Ret$ for the relative variable $\Delta N^K_{R}$ investigated by considering the linear model 
\begin{equation}
\label{linearDNR}
\widehat{\Delta}N^K_{R} (t)=\alpha_{S_{R}} \widehat{S}_{R} (t)+ \alpha_{Ret} \widehat{R}et (t)+\epsilon(t)
\end{equation}
where $\widehat{\Delta}N^K_{R}$, $\widehat{S}_{R}$, and $\widehat{R}et$ are standardized versions with zero mean and unitary variance of $\Delta N^K_{R}$, $S_{R}$, and $Ret$, respectively. By comparing the results of Table  \ref{DNRCtg} with the results reported in Table \ref{DNAbCtg}  we conclude that the observations done for $\widehat{\Delta}N^K_{A}$, $\widehat{R}et$ and $\widehat{S}_{A}$ are very similar to the ones obtained for the relative variables $\widehat{\Delta}N^K_{R}$, $\widehat{R}et$ and $\widehat{S}_{R}$, and therefore do not depend significantly on the specific definition of the sentiment indicator. The only difference we detect is observed for companies and households. It concerns the values of the coefficient  $\alpha_{S_{R}} $ which are not consistent with zero within the 5\%-95\% confidence interval.

The partial correlations shown in Table \ref{DNAbCtg} and \ref{DNRCtg} indicate that $|\rho(N^K_A,S_A|Ret)|$ $\ll$ $|\rho(N^K_A,Ret|S_A)|$ and $|\rho(N^K_R,S_R|Ret)|$ $\ll$ $|\rho(N^K_R,Ret|S_R)|$ for all the categories of investors. The highest values of partial correlations, $\rho(N^K_A,S_A|Ret)$ and $\rho(N^K_R,S_R|Ret)$, are observed for companies and households. However, in general, these correlations are quite small in absolute value (between $5\%$ and $8\%$), and very close to the noise level. For the other categories, these partial correlations are even smaller, and statistically not significant. 

The joint analysis of absolute and relative imbalance between buyers and sellers leads us to the following conclusions. The activity of governmental and non profit organizations is very poorly explained by return and news sentiment. Of the two factors, return plays clearly a major role. Households and companies are those for which sentiment and returns have the best explanatory power of their trading action. Return is clearly more important, but sentiment has also some explanatory power, especially when one consider the relative imbalance between buyers and sellers. For financial and foreign organizations the variance explained by the regressions is somewhat intermediate between the two pairs of categories above, but in general returns have a much higher explanatory power and sentiment plays a negligible role.

In conclusion, the regression analysis shows that the total flux of news is significantly correlated with the decision to trade.
This seems to indicate that, on a daily time scale, news move investors to trade.  However, we find that most of the times the sentiment indicator is not significantly correlated with the imbalance between buyers and sellers. One possible explanation is that the sentiment indicator does not discriminate accurately a good news  from bad news. As mentioned above, sentiment indicator has a correlation with returns of value around $0.15$ and therefore it seems possible that a more accurate semantic indicator could better capture how good or bad is the news. Another possibility is that most of the times investors do not agree on the positivity or negativity of the news and/or that they react in an heterogeneous way to the same news.

\section{Conclusions}
\label{Conclusions}
By using a linear regression model to describe the number of investors trading the Nokia stock as a linear combination of a proxy of news and a proxy of stock volatility, we have shown that  the trading activity of all the investigated categories of investors is significantly correlated with both the flux of news and the daily volatility. The relative role of exogenous (news) and endogenous (volatility) factors is, in general, difficult to disentangle because the two variables are highly correlated. By assuming a linear model, we have shown that the relative relevance of the two variables changes across the different categories of investors. The dependency from volatility turned out to be more pronounced than the one from news, for companies, financial institutions, households, and foreign organizations.

The second conclusion concerns the relationship between the sentiment of news, Nokia return, and the market polarization towards buying or selling for the different investors' categories. Our results show that the sentiment time series is correlated with Nokia return. We have also shown that both the Nokia return and the sentiment of news explain part of the trading polarization dynamics of some categories, according to a linear model. Specifically, we observed that all categories have a contrarian like response to return, and that companies and households have a positive, although small sized, reaction to positive sentiment of news.\\

{\bf Acknowledgments} Authors acknowledge Thomson Reuters for providing the NewScope archive data and Euroclear for granting access to the ownership data of single investors. FL acknowledges partial support by the grant SNS11LILLB ``Price formation, agent's heterogeneity, and market efficiency". JP acknowledges Magnus Ehrnrooth Foundation, and Jenny and Antti Wihuri Foundation for financial support.

 \appendix
 \section{Proof of Eq.~(\ref{lineartopartial})}
\label{lintopart}

Suppose we aim at explaining the behavior of a stochastic variable $y$ with time samples $\{y_1,...,y_T\}$ as a linear combination of two correlated variables $x_1$ and $x_2$ with sample series $\{x_{1,1},...,x_{1,T}\}$ e $\{x_{2,1},...,x_{2,T}\}$, respectively. Synchronous sampling of the three variables is assumed. We focus on the linear model:
\begin{equation}
\label{model}
y=\alpha_1 \cdot x_1+ \alpha_2 \cdot x_2 + \beta \epsilon,
\end{equation}
where $\epsilon$ is the idiosyncratic term. If we assume (without loss of generality) that all the variables, including $y$, are standardized to zero mean and unit variance, then the value of the coefficient $\beta$ is determined by the following equation for the variance:
\begin{equation}
\label{detbeta}
<y^2>=1=\alpha_1^2+\alpha_2^2 +2 \alpha_1 \alpha_2 \rho_{1,2}+\beta^2,
\end{equation}
where $\rho_{1,2}$ is the correlation between $x_1$ and $x_2$. We shall calculate the value of $\beta$ by using this equation afterwards, that is once the value of $\alpha_1$ and $\alpha_2$ will be regressed from data. According to the least squares method, estimates of these two parameters are obtained by minimizing the following function:
\begin{equation}
\label{minquad}
f(\alpha_1,\alpha_2)=\sum_{i=1}^T\left(y_i-\alpha_1\cdot x_{1,i}-\alpha_2\cdot x_{2,i}\right)^2.
\end{equation}
We have
\begin{align}
\label{deralpha1}
0=\frac{\partial f(\alpha_1,\alpha_2)}{\partial \alpha_1}=-2 T\cdot \frac{1}{T}\sum_{i=1}^T\left(y_i-\alpha_1\cdot x_{1,i}-\alpha_2\cdot x_{2,i}\right)\cdot x_{1,i}= \nonumber \\
-2 T\cdot \left(\rho_{1,y}- \alpha_1 -\alpha_2 \rho_{1,2}\right ),
\end{align}
where $\rho_{1,y}$ and $\rho_{1,2}$ are the estimated correlations between $x_1$ and $y$ and between $x_1$ and $x_2$, respectively. In the previous equation, the last equality is based on the assumption that data has also been standardized. Similarly, by differentiating with respect to $\alpha_2$ we obtain:
\begin{align}
\label{deralpha2}
0=\frac{\partial f(\alpha_1,\alpha_2)}{\partial \alpha_2}=-2 T\cdot \frac{1}{T}\sum_{i=1}^T\left(y_i-\alpha_1\cdot x_{1,i}-\alpha_2\cdot x_{2,i}\right)\cdot x_{2,i}=\nonumber \\ 
-2 T\cdot \left(\rho_{2,y}- \alpha_1 \rho_{1,2} -\alpha_2\right ),
\end{align}
where $\rho_{2,y}$ is the estimated correlation between $x_2$ and $y$. The latter two equations allow one to estimate $\alpha_1$ and $\alpha_2$ as
\begin{eqnarray}
\label{alpha12}
\alpha_1=\frac{\rho_{1,y}-\rho_{2,y}\cdot \rho_{1,2}}{1-\rho_{1,2}^2},\nonumber \\
\alpha_2=\frac{\rho_{2,y}-\rho_{1,y}\cdot \rho_{1,2}}{1-\rho_{1,2}^2}.
\end{eqnarray}
Finally, these results are used in Eq. (\ref{detbeta}) to estimate parameter $\beta$:
\begin{align}
\label{detbetafinal}
\beta^2=&1-\alpha_1^2 - \alpha_2^2 - 2 \alpha_1 \alpha_2 \rho_{1,2}=&\nonumber \\
&\frac{1-\rho_{1,2}^2-\rho_{1,y}^2-\rho_{2,y}^2+2 \rho_{1,2}\, \rho_{1,y} \, \rho_{2,y}}{1-\rho_{1,2}^2}=\frac{|\Gamma|}{1-\rho_{1,2}^2},
\end{align}
where $|\Gamma|$ is the determinant of the correlation matrix of the three variables $x_1$, $x_2$, and $y$. This result indicates that $\beta^2$ is always non-negative (as it should be), and it represents the fraction of variance of $y$ that is not explained by the two variables $x_1$ and $x_2$, as it is in the standard linear models based on independent variables. The case of uncorrelated $x_1$ and $x_2$, is simply obtained by setting $\rho_{1,2}=0$ in the previous equations. The result is that $\alpha_1=\rho_{1,y}$, $\alpha_2=\rho_{2,y}$ and $\beta^2=1-\rho_{1,y}^2-\rho_{2,y}^2$, as expected.\\
Hence, the estimates of $\alpha_1$ and $\alpha_2$ are related to the partial correlations
\begin{eqnarray}
\label{partialcorr}
\rho(y,x_1|x_2)=\frac{\rho_{1,y}-\rho_{2,y}\cdot \rho_{1,2}}{\sqrt{(1-\rho_{2,y}^2)\cdot(1-\rho_{1,2}^2)}},\nonumber \\
\rho(y,x_2|x_1)=\frac{\rho_{2,y}-\rho_{1,y}\cdot \rho_{1,2}}{\sqrt{(1-\rho_{1,y}^2)\cdot(1-\rho_{1,2}^2)}}
\end{eqnarray}
through the equations:
\begin{eqnarray}
\label{alphatopartial}
\alpha_1=\frac{\sqrt{(1-\rho_{2,y}^2)\cdot(1-\rho_{1,2}^2)}}{1-\rho_{1,2}^2} \cdot \rho(y,x_1|x_2),\nonumber\\
\alpha_2=\frac{\sqrt{(1-\rho_{1,y}^2)\cdot(1-\rho_{1,2}^2)}}{1-\rho_{1,2}^2} \cdot \rho(y,x_2|x_1).
\end{eqnarray}
In summary:
\begin{equation}
\label{alphatopartial2}
\frac{\alpha_1}{\alpha_2}=\frac{\rho(y,x_1|x_2)}{\rho(y,x_2|x_1)}\cdot \sqrt{\frac{1-\rho_{2,y}^2}{1-\rho_{1,y}^2}}.
\end{equation}
which is equivalent to Eq.(9).

\bibliographystyle{model5-names}
\bibliography{Biblio_JBF}

\end{document}